# Quantum Random Synthetic Skyrmion Texture Generation, a Qiskit Simulation

Hillol Biswas

Department of Electrical and Computer Engineering

Democritus University of Thrace

XANTHI, Greece

## Abstract

An integer winding, i.e., topological charge, is a characteristic of skyrmions, which are topologically nontrivial spin patterns in magnets. They emerge when smooth two-dimensional spin configurations are stabilized by conflicting interactions such as exchange, anisotropy, the Dzyaloshinskii-Moriya interaction, or geometric frustration. These nanoscale textures, which are typically a few to tens of nanometers in size, are strong 'particle-like' excitations because they are shielded by energy barriers connected to their topology. By exploiting their helicity, i.e., spin rotation angle or associated internal modes, as a two-level system, skyrmions can function as quantum bits or qubits. Two quantized helicity states of a nanometer-scale skyrmion encode the logical value states in a 'skyrmion qubit.' Interestingly, skyrmion qubits are topologically protected and macroscopic, i.e., they involve a large number of spins; however, external influences can still affect them. When the texture is tiny and disconnected, the helicity angle of the skyrmion becomes quantized. A qubit basis is made up of the lowest two energy eigenstates, i.e., symmetric or antisymmetric superpositions of opposite helicity, for example. Therefore, Skyrmion textures can provide valuable insights for different purposes. However, is it possible to synthetically generate skyrmion textures using quantum computing? This paper investigates the possibility and generates a few hundred different textures, producing sample comparisons from various types, which indicate a novel direction for skyrmion-based research based on quantum randomness and other criteria.

## Introduction

Magnetic skyrmions are particle-like spin patterns at the nanoscale that are stabilized by topological protection and chiral interactions (Dzyaloshinskii–Moriya) . They are extremely stable yet mobile information bits because of their integer "winding" charge, which provides high energy barriers against degradation [1],[2]. Skyrmions are powered by incredibly low current densities (~$10^6$ A/m²) and can be as small as a few nanometers. With spin torques or fields, a skyrmion can be produced, erased, and shifted. In practice, it encodes a "1" (spin swirl) vs. a "0" (background). Skyrmions are appealing for nonvolatile spintronic devices because of their compact size, low depinning currents, and high density. Their nontrivial spin topology also offers the possibility of additional functionality [2][3].

• Important characteristics include: incredibly small lateral size (nm scale); topologically stable whirl-like spin configurations (Bloch/Néel type); and core magnetization opposite the periphery [2].
• Mobility: Powered by spin-orbit or spin-transfer torques, usually at currents significantly lower than those required for domain walls [2][1].
• Information encoding: A bit is represented by the presence or absence of a skyrmion. Local currents or fields have the ability to write or erase them, and magnetoresistive readout can detect them.
• Stability: Long lives are possible even in warm conditions because they are shielded from their topological charge by energy barriers [2].
These characteristics have led to ideas for spintronic memory and logic based on skyrmions. Racetrack memory architectures, in-memory computation, and neuromorphic components are all naturally suited to skyrmions. Furthermore, their quantized collective excitations (helicity, for example) provide fascinating new directions in quantum information.

All-magnetic logic can alternatively be based on skyrmions. Different proposals use controlled annihilations, conversions, and skyrmion collisions to build logic gates. In order to enable basic logic operations, Zhang et al. (2015) demonstrated theoretically that solitary skyrmions can be transformed, replicated, or combined (even into other textures like bimerons) within predefined nanostructures [4]. They guided skyrmions into junctions where they combine or annihilate in order to demonstrate logic gates like AND and OR. Recently, fully reconfigurable skyrmion logic families have been proposed. One approach uses patterned Dzyaloshinskii–Moriya (DMI) "chirality barriers" to selectively pin skyrmions on a single nanotrack. According to Yu et al. (2022), a whole set of Boolean gates (AND, OR, NOT, NAND, NOR, XOR, and XNOR) can be implemented on-demand by reversing these barriers [5]. Depending on the chirality profile of the junction, skyrmions that contact it either pass, merge, or annihilate. These skyrmion logic gates would be cascadeable and non-volatile.

Researchers are looking into skyrmions for quantum information in addition to classical spintronics. A number of possibilities are made possible by the quantum dynamics of microscopic skyrmions: Majorana modes can be produced by hybridizing with superconductors, and their collective excitations can act as quantum two-level systems. These potential are described in depth in recent theoretical and review articles. In their discussion of skyrmion qubits, Psaroudaki et al. (2024) point out that nano-skyrmions possess quantized helicity states that are capable of tunneling across orientations, creating a macroscopic qubit with topological protection [6]. In practice, a 0/1 might be encoded by adjusting the helicity of the skyrmion using spin torques or manufactured anisotropy. By bridging the gap between spintronics and quantum technologies, Petrović et al.

(2025) highlight that magnetic skyrmions are among the smallest magnetically-ordered textures and are hence excellent candidates to display coherent quantum behavior [2].

Topological quantum computing is a further method that uses skyrmions to create topological superconductivity. It has been suggested that a ferromagnet's skyrmion can bind Majorana zero modes at its core when it is proximitized by an s-wave superconductor. The holy grail of fault-tolerant qubits, non-Abelian statistics, would be obeyed by such skyrmion–vortex pair states. In fact, theoretical investigations have demonstrated that nontrivial braiding of Majorana bound states (MBS) is possible in hybrid structures containing skyrmions [6]. The development of magnetic skyrmion lattices under superconducting films is an example of an experimental advancement that is only getting started. In conclusion, skyrmions are being investigated as topological qubits in two ways: (i) as platforms for non-Abelian anyons (Majorana states) when interfaced with superconductivity, and (ii) as quantum two-level systems through their helicity excitations [6]. Although real quantum devices are still a ways off, Petrović et al. point out that these advancements generate "new paradigms" that combine spintronics and quantum computing [2].

# Background

Since the introduction of quantum computing through stalwarts like Deutsch[7], Shor[8], [9], Bernstein and Vazirani[10], Grover[11], and Simon [12] algorithms, Quantum research in applications of different domains is fast spreading. To find lightweight dark-matter candidates (axions/dark photons), quantum sensors such as atomic magnetometers, superconducting qubits, microwave cavities, and long-baseline sensor networks are now being used. These methods increase sensitivity beyond conventional bounds by utilizing correlated sensor networks and quantum enhancement [13]. To speed up and increase the sensitivity of searches, superconducting qubits and other quantum readout/modalities are being proposed (e.g., enabling single-photon-level readout in haloscopes or increasing axion scan speeds) [14]. In contrast to qubit discretizations, new formulations that transfer gauge fields and continuum degrees of freedom to continuous-variable (CV) quantum hardware (bosonic modes) may offer more natural encodings for field theories [15]. To evaluate theoretical models in controlled environments, analog and digital quantum simulators are used to simulate intricate many-body processes that are pertinent to condensed matter and high-energy systems, such as pair creation, topological phases, and thermalization [16].

At least nine different skyrmionic textures in model magnetic systems have been effectively categorized in recent works using sophisticated simulations and machine learning; more could be added as experimental methods improve [17].

Depending on the material properties and external fields, skyrmion textures might appear as stripes, lattices, single creatures, or even as intricate, linked, or periodic arrangements [18].

There are reports of novel topologies involving vorticity, polarization, and three-dimensional linkages, indicating an almost limitless field for further research [19], [20].

There is no limit to the number of known skyrmion textures; as scientists investigate novel materials, dimensionalities, and magnetic configurations, the field is rapidly growing and new textures are frequently reported [21], [22].

Therefore, Skyrmionics is a fast developing field. Proof-of-concept demonstrations are giving way to hybrid systems and more reliable device concepts in the current effort. Reducing device variability, lowering operating currents, and incorporating skyrmion devices into traditional spintronic designs are areas of research on the technological front. At the same time, state-of-the-art research is investigating quantum skyrmionics in an effort to use the quantized degrees of freedom of skyrmions for topological computing and quantum bits. Skyrmions present "opportunities to merge the fields of spintronics, quantum computing, and strongly correlated systems," as Petrović et al. highlight [2]. Skyrmion-based technologies have the potential to provide ultra-dense memory, innovative logic/neuromorphic processors, and possibly the foundation for fault-tolerant quantum computers if these obstacles can be overcome.

# Quantum Computing-Based Skyrmion Texture Generation

Quantum computing-based fractal image generation [23] renders different iterative synthetic fractal images. Based on the same principle, when tailor-made with skyrmion images with spin fields, it is possible to create synthetic varieties of images. Fig. 1 depicts the steps of the flow-chart for batches of different skyrmion images iteratively. Fig. 2 depicts the quantum circuit prepared with six qubits for the generation of synthetic skyrmion images.

Qiskit [24], [25] It is a software development kit (SDK) that is widely used for quantum computing purposes. Using Qisit and the corresponding library, this paper's contribution is the synthetic generation of different Skyrmion-textured images as a batch through a quantum computing technique. Mimicking the Skyrmion texture of the images for potential novel applications is itself the novelty of the paper. Nonetheless, this paper demonstrates that large-scale generation of synthetic skyrmion images is plausible.

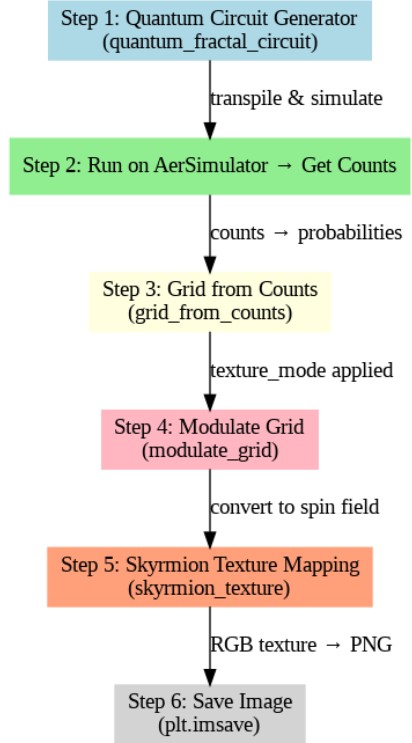

Fig. 1: Flow chart of Skyrmion Synthetic Texture Generation Process through Quantum Computing

# Results:

Based on the approach as elicited above, the quantum circuits prepared, Fig. 2 depicts the different gates viz. CNOT used for building the circuits with six qubits and depth of six.

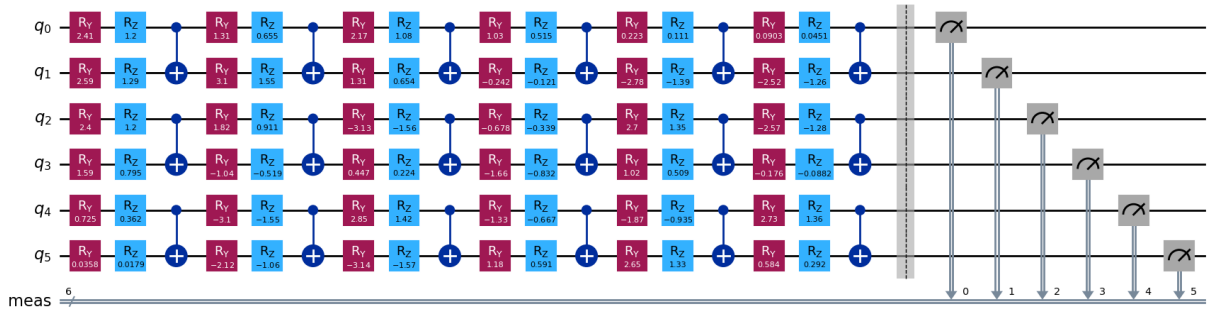

Fig. 2: Quantum Circuit for Skyrmion Texture Generation

The general form of state vector of a quantum circuit is given by Eq. (1):

$$|\psi\rangle = \sum_{i=0}^{2^n - 1} c_i |i\rangle \qquad (1)$$

Where n is the number of qubits, $|i\rangle$ are the computational basis states and $c_i$ are complex amplitudes for each state. Fig. 3 encodes the superposition of all the sixty-four states.

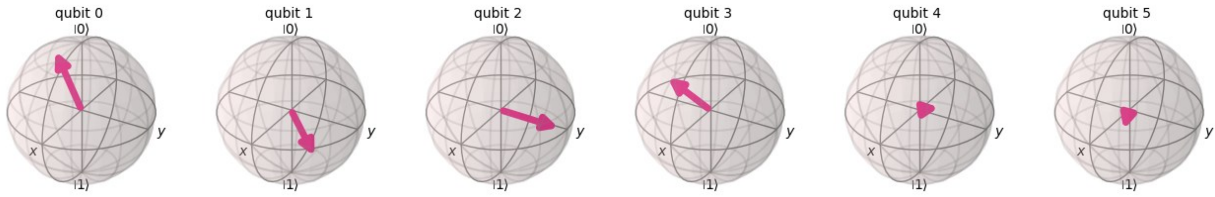

*Fig. 3: Six Qubit Multi-vector plot*

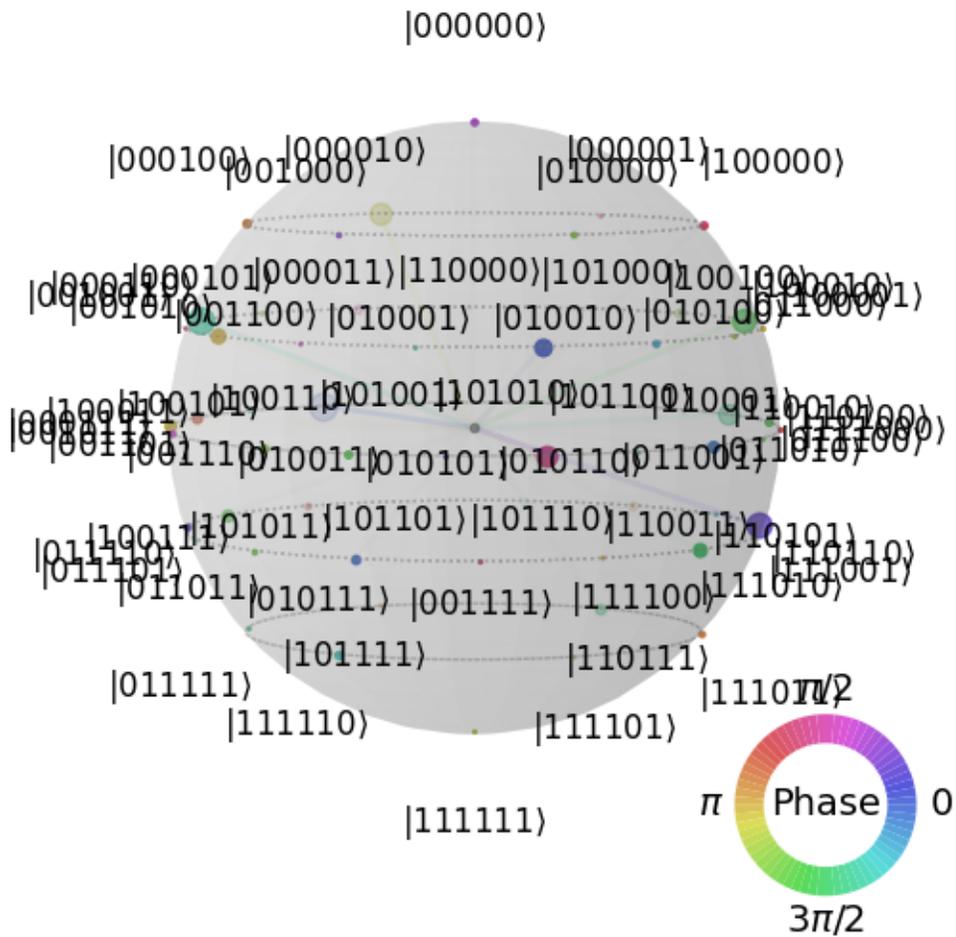

*Fig. 4: QSphere Plot of the Skyrmion Quantum Circuit*

Fig. 4 depicts the Qsphere plot of the quantum circuit comprising six qubits. The quantum circuit-based synthetic generation created four types of skyrmion-textured images, each with fifty images using the same seed and depth.

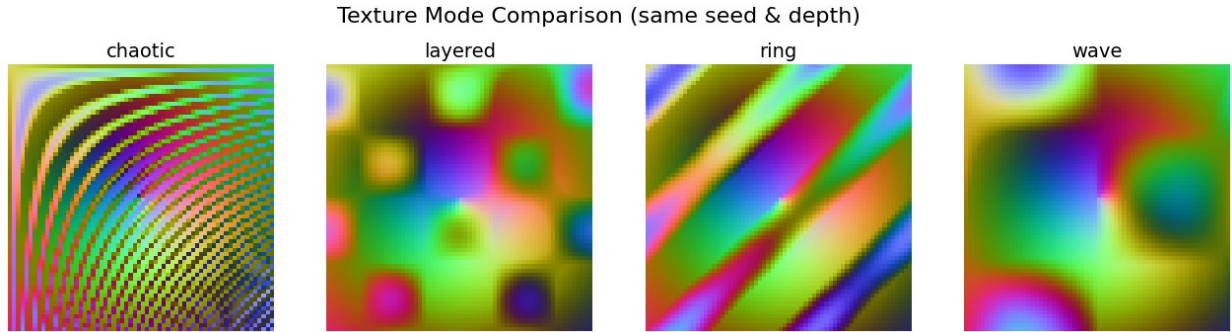

*Fig. 5: Synthetic Skyrmion Texture Generation – Chaotic, Layered, Ring and Wave Type*

Fig. 5, using the same seed and depth settings but distinct algorithms for texture formation, this graphic compares various texture production modes. Each sub-image depicts the following chaotic, layered, ring and wave types. Chaotic type creates incredibly erratic, cacophonous, and jagged interference-like patterns. The structure is visually complicated, resembling overlapping wavefronts with high-frequency fluctuations. Layered textures are smoother and more distinct than chaotic ones, resembling stacked or overlapping blobs or patches. It highlights areas of color that resemble blocks. Ring type creates structures that resemble bands or slanted stripes with a recurring circular or elliptical pattern. The wave type makes color gradients flow across the image like waves, creating smoother, undulating transitions.

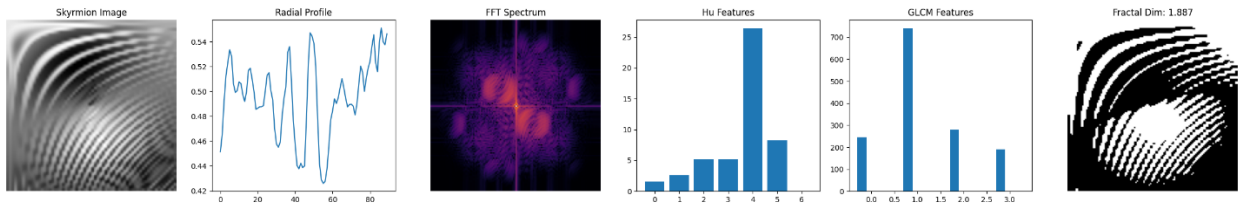

*Fig. 6: A Sample Chaotic Skyrmion Texture and its characteristics*

Each image attributes are separately identified through radial profile, fast Fourier transform, Hu features, GLCM features, and fractal dimensions. A Sample of each of the above types of synthetically generated images is revealed by Fig. 6 – 9 with different characteristics. The FFT spectrum reveals each distinct category and its corresponding fractal dimensions, which are 1.887, 1.829, 1.832, and 1.857, as subtle differences are exhibited in the images.

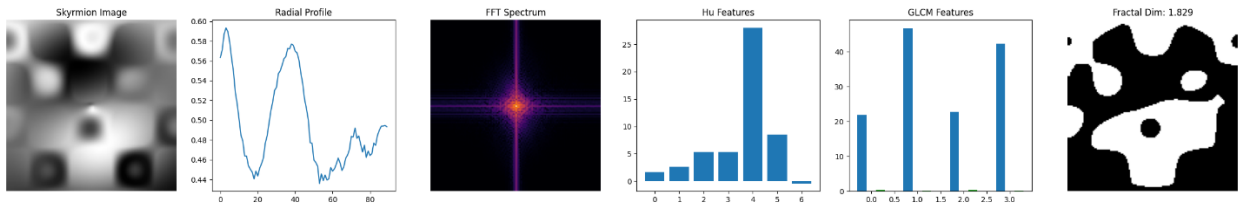

*Fig. 7: A Sample Layered Skyrmion Texture and its characteristics*

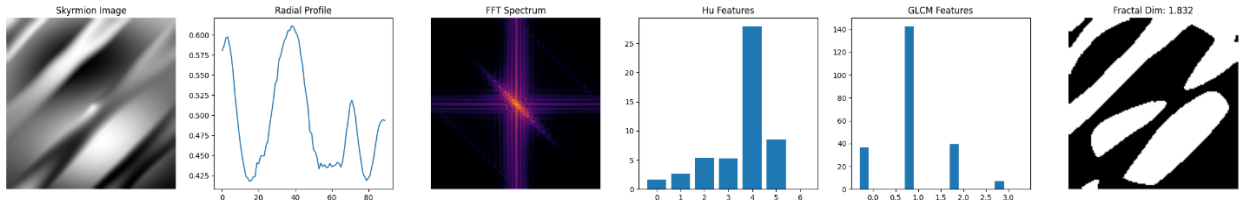

*Fig. 8: A Sample Ring Skyrmion Texture and its characteristics*

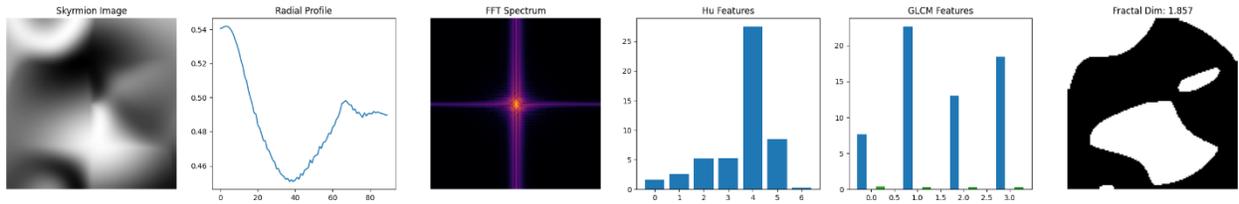

*Fig. 9: A Sample Wave Skyrmion Texture and its characteristics*

Further, the alternate attributes of the same images have been captured through pixel histogram, canny edge detection, 2D auto-correlation, wavelet approximation, and local binary patterns, Fig. 10-13 of the corresponding sample images of Fig. 2 and Fig. 6-9 as described above.

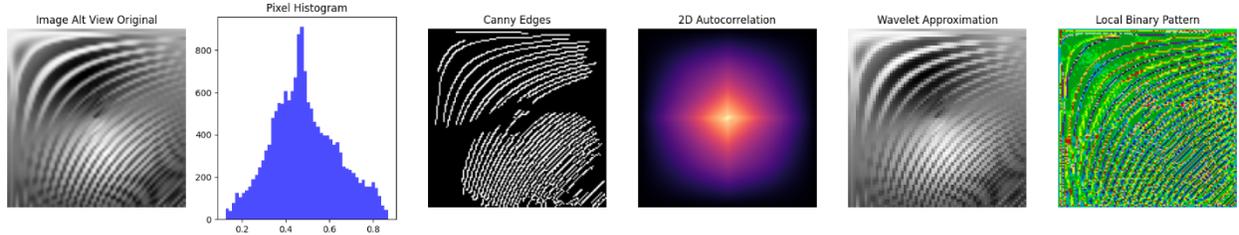

*Fig. 10: Display Alternate Image Characteristics of Chaotic Sample*

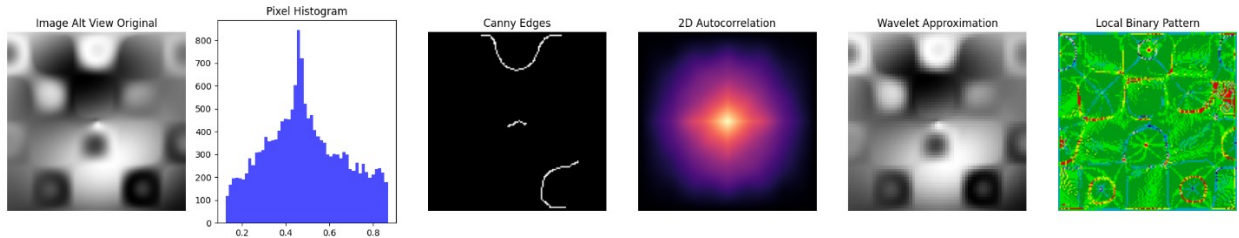

*Fig. 11: Display Alternate Image Characteristics of Layered Sample*

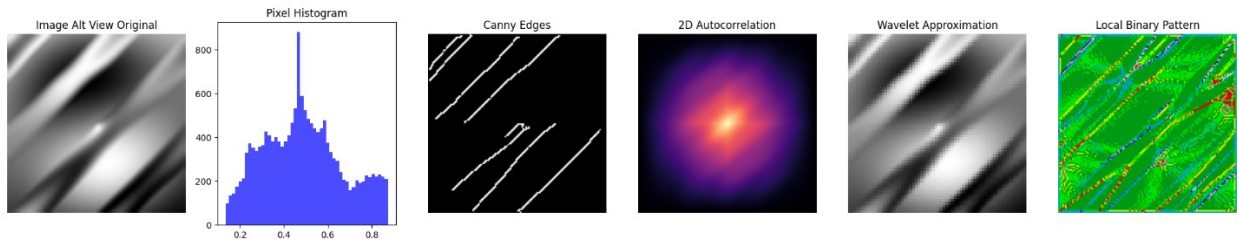

*Fig. 12: Display Alternate Image Characteristics of Ring Sample*

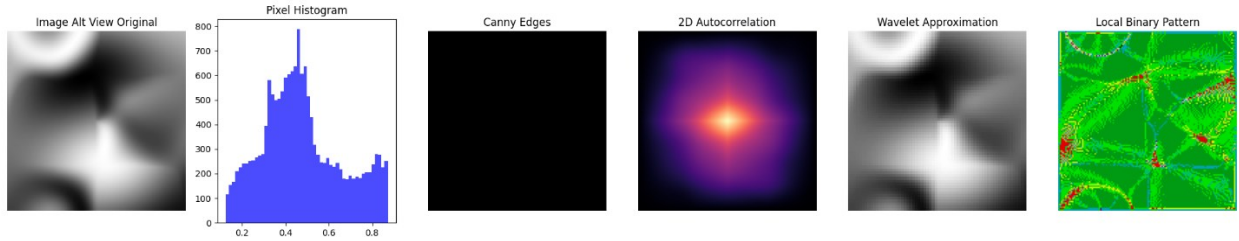

*Fig. 13: Display Alternate Image Characteristics of Wave Sample*

Richer edges, greater LBP complexity, and directional autocorrelation are the results of the repeating curving lines and delicate, high-frequency texture in Fig. 10. Coarse, low-frequency blobs with fewer edges, a simpler wavelet approximation, isotropic correlation, and a lower LBP complexity predominate in Fig. 11. Fig. 12 denotes a Blob-based image with low structural orientation, circular autocorrelation, weak edges, and isotropic features. Anisotropic features, firm directional edges, elongated autocorrelation, and orientation preservation are all present in the stripe-based image shown.

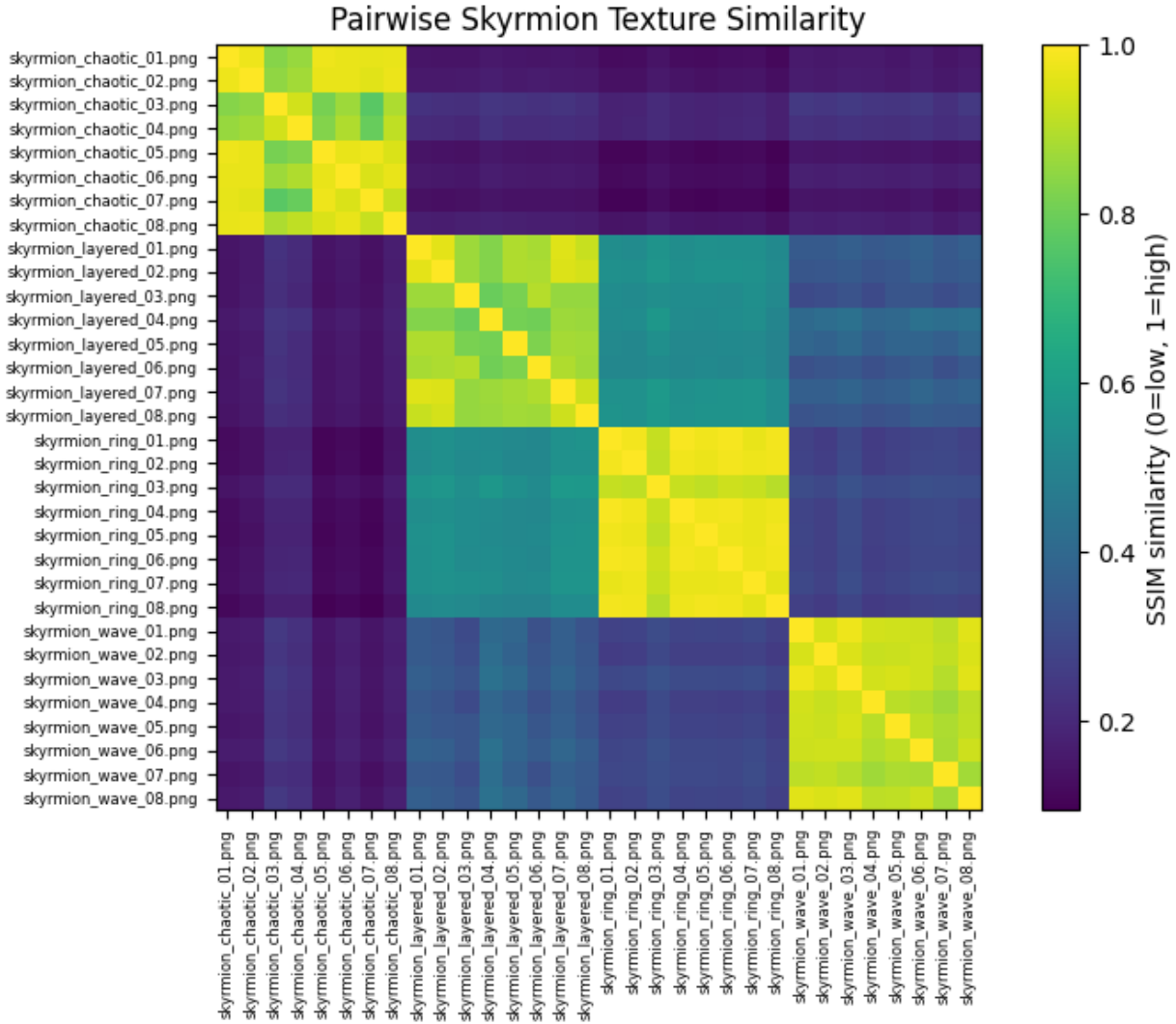

*Fig. 14: Pairwise Skyrmion Texture Similarity*

A popular technique for assessing how similar (or different) two photographs are is called the Structural Similarity Index Measure (SSIM). It was first presented as an enhancement over the pixel-by-pixel error-only metrics of PSNR (Peak Signal-to-Noise Ratio) and MSE (Mean Squared Error). SSIM examines perceived structural information rather than raw differences, which is more in line with how people assess image quality. Fig. 14 compares the heatmap of the four types of images produced for the first eight numbers with their pairwise similarities.

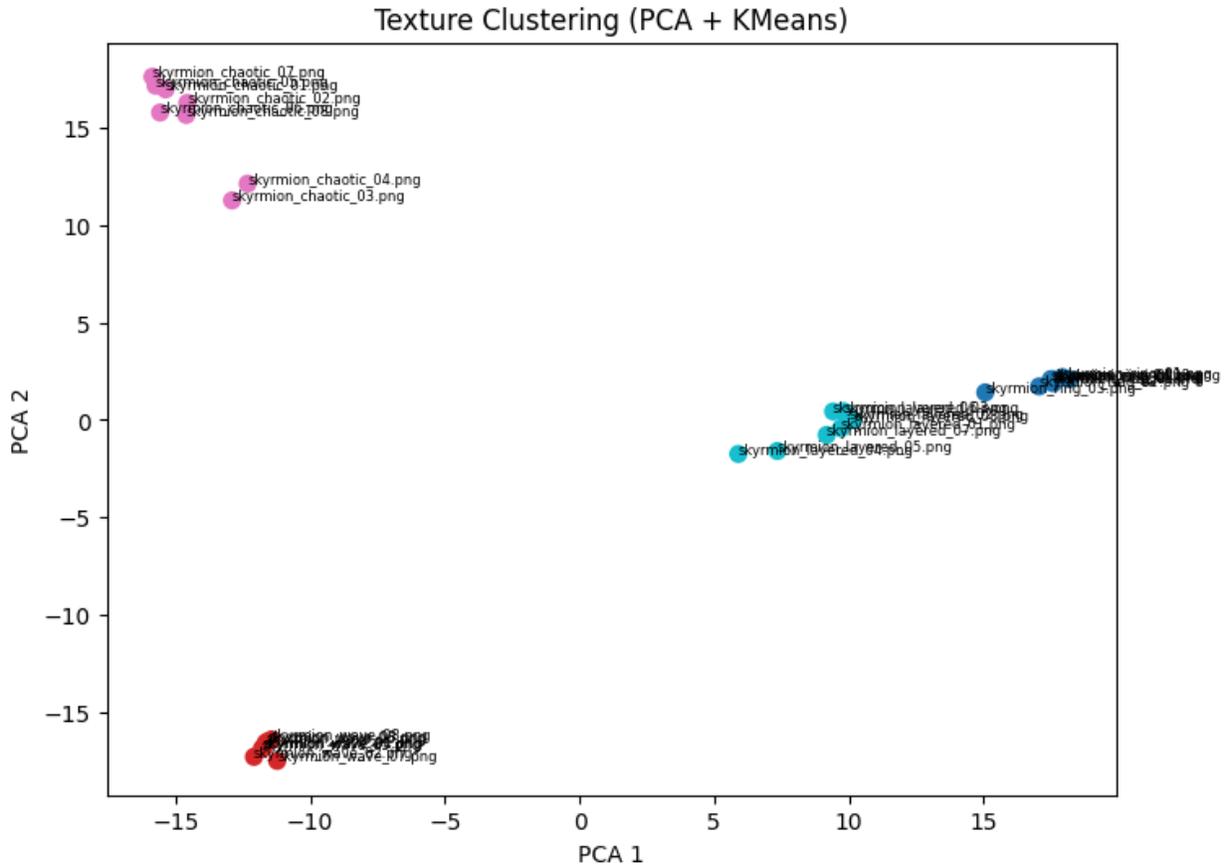

*Fig. 15: Embedding and Clustering on two principal components*

Fig. 15 denotes the two principal component based clustering of the each eight type of images which exhibits different clustering revealing each image are unique and different as synthetically generated.

# Discussion and Conclusion:

Collectively, these Fig. show how feature extraction pipelines differentiate between periodic radial, anisotropic, and isotropic textures. Texture categorization, material analysis, medical imaging (tissue textures), and surface patterns all benefit greatly from this type of analysis. The comparison also demonstrates how various structural aspects are highlighted by the same feature extraction workflow (histogram, edges, autocorrelation, wavelets, LBP). In computer vision/ image analysis, and texture classification, these properties are essential because they can differentiate between smooth (isotropic) and structured (anisotropic) textures.

Because of its high mobility and absence of the skyrmion Hall effect, synthetic antiferromagnetic (SAF) skyrmions are becoming new information carriers. It is still difficult to differentiate SAF skyrmions from their ferromagnetic counterparts using imaging methods like magneto-optical microscopy [26], [27]. Attributed to characteristics like high mobility and a suppressed skyrmion Hall effect, new synthetic skyrmion textures can offer novel functionality for memory, logic, and magnonic applications vi. Spintronic devices. Finding skyrmion types with unique and practical properties for theoretical and experimental investigation is more likely when the material candidates and design approaches are expanded. In conclusion, controlled creation of artificial skyrmion textures is not only feasible but also a promising avenue for the development of next-generation magnetic technologies and the identification of novel skyrmion varieties.

This work demonstrates that synthetically quantum computing approach based on skyrmion images are possible to generate and the texture patterns are comparable to each other for finding out the subtle difference using spin texture by different comparison matrices. The study is limited to generation of few hundred images however, is has the potential of generating significant numbers which may also be possible to check with large dataset principles and techniques.

A novel platform at the intersection of quantum information and spintronics is provided by skyrmions. They are appealing for quantum devices due to their topological nature and electric/magnetic controllability. Skyrmionics is a fast-developing field. Proof-of-concept demonstrations are giving way to hybrid systems and more reliable device concepts in the current effort. Reducing device variability, lowering operating currents, and incorporating skyrmion devices into traditional spintronic designs are areas of research on the technological front. At the same time, state-of-the-art research is investigating quantum skyrmionics in an effort to use the quantized degrees of freedom of skyrmions for topological computing and quantum bits.

# References:

[1]  S. Luo and L. You, "Skyrmion devices for memory and logic applications," *APL Mater.*, vol. 9, no. 5, 2021, doi: 10.1063/5.0042917.

[2]  A. P. Petroviˊcpetroviˊc, C. Psaroudaki, P. Fischer, M. Garst, and C. Panagopoulos, "Colloquium: Quantum Properties and Functionalities of Magnetic Skyrmions," 2024.

[3]  M. Zhao *et al.*, "Electrical detection of mobile skyrmions with 100% tunneling magnetoresistance in a racetrack-like device," *npj Quantum Mater.*, vol. 9, no. 1, p. 50, 2024, doi: 10.1038/s41535-024-00655-1.

[4]  X. Zhang, M. Ezawa, and Y. Zhou, "Magnetic skyrmion logic gates: conversion, duplication and merging of skyrmions," *Sci. Rep.*, vol. 5, no. 1, p. 9400, 2015, doi: 10.1038/srep09400.

[5]  D. Yu, H. Yang, M. Chshiev, and A. Fert, "Skyrmions-based logic gates in one single nanotrack completely reconstructed via chirality barrier," *Natl. Sci. Rev.*, vol. 9, no. 12, 2022, doi: 10.1093/nsr/nwac021.

[6]  C. Psaroudaki, E. Peraticos, and C. Panagopoulos, "Skyrmion Qubits: Challenges For Future Quantum Computing Applications".

[7]  D. Deutsch, "Quantum Theory, the Church-Turing Principle and the Universal Quantum Computer," *Proc. R. Soc. Lond. A. Math. Phys. Sci.*, vol. 400, no. No. 1818(Jul. 8, 1985), pp. 97–117, 1985, [Online].

Available: https://www.cs.princeton.edu/courses/archive/fall06/cos576/papers/deutsch85.pdf

[8]     P. W. Shor, "Polynomial-Time Algorithms for Prime Factorization and Discrete Logarithms on a Quantum Computer," vol. 41, no. 2, pp. 303–332, 1999.

[9]     P. W. Shor, "Quantum Computing," vol. ICM, 1998.

[10]    E. Bernstein and U. Vaziranit, "Quantum complexity theory," *Proc. Annu. ACM Symp. Theory Comput.*, vol. Part F1295, pp. 11–20, 1993, doi: 10.1145/167088.167097.

[11]    L. K. Grover, "A fast quantum mechanical algorithm for database search," pp. 212–219, 1996, doi: 10.1145/237814.237866.

[12]    D. R. Simon, "On the power of quantum computation," *SIAM J. Comput.*, vol. 26, no. 5, pp. 1474–1483, 1997, doi: 10.1137/S0097539796298637.

[13]    M. Jiang *et al.*, "Long-baseline quantum sensor network as dark matter haloscope," *Nat. Commun.*, vol. 15, no. 1, p. 3331, 2024, doi: 10.1038/s41467-024-47566-0.

[14]    C. Braggio *et al.*, "Quantum-Enhanced Sensing of Axion Dark Matter with a Transmon-Based Single Microwave Photon Counter," *Phys. Rev. X*, vol. 15, no. 2, p. 21031, Apr. 2025, doi: 10.1103/PhysRevX.15.021031.

[15]    G. Adesso, A. Serafini, and F. Illuminati, "Quantum computation of SU(2) lattice gauge theory With Continuous Variables," vol. 06, 2025.

[16]    "Quantum simulators in high-energy physics – CERN Courier."

[17]    D. Feng, Z. Guan, X. Wu, Y. Wu, and C. Song, "Classification of skyrmionic textures and extraction of Hamiltonian parameters via machine learning," *Phys. Rev. Appl.*, vol. 21, no. 3, p. 034009, Mar. 2024, doi: 10.1103/PHYSREVAPPLIED.21.034009/FIGURES/5/THUMBNAIL.

[18]    H. Wu, X. Hu, K. Jing, and X. R. Wang, "Size and profile of skyrmions in skyrmion crystals," *Commun. Phys. 2021 41*, vol. 4, no. 1, pp. 1–7, 2021, doi: 10.1038/s42005-021-00716-y.

[19]    X. Yao and S. Dong, "Vector vorticity of skyrmionic texture: An internal degree of freedom tunable by magnetic field," *Phys. Rev. B*, vol. 105, no. 1, p. 14444, 2022, doi: 10.1103/PHYSREVB.105.014444/FIGURES/7/THUMBNAIL.

[20]    Y. Shen, Q. Zhang, P. Shi, L. Du, X. Yuan, and A. V Zayats, "Optical skyrmions and other topological quasiparticles of light," 2024.

[21]    S. Koraltan *et al.*, "Signatures of higher order skyrmionic textures revealed by magnetic force microscopy," 2025.

[22]    D. Singh *et al.*, "Transition between distinct hybrid skyrmion textures through their hexagonal-to-square crystal transformation in a polar magnet," *Nat. Commun.*, vol. 14, no. 1, pp. 1–12, 2023, doi: 10.1038/S41467-023-43814-X;TECHMETA=120,128;SUBJMETA=119,301,639,766,997;KWRD=MAGNETIC+PROPERTIES+AND+MATERIALS.

[23]    H. Biswas, "Quantum-Circuit-Based Visual Fractal Image Generation in Qiskit and Analytics," 2025.

[24]    "Qiskit | IBM Quantum Computing." Accessed: Aug. 13, 2024. [Online]. Available: https://www.ibm.com/quantum/qiskit

[25]    Qiskit Team, "Release News: Qiskit SDK v2.0 is here!"

[26]    S. Zhang, A. K. Petford-Long, and C. Phatak, "Creation of artificial skyrmions and antiskyrmions by anisotropy engineering," *Sci. Rep.*, vol. 6, no. 1, pp. 1–10, 2016, doi: 10.1038/SREP31248;TECHMETA=128;SUBJMETA=119,301,357,639,997;KWRD=MAGNETIC+PROPERTIES+AND+MATERIALS.
Page 13 of 14

[27] Y. Zhao *et al.*, "Identifying and Exploring Synthetic Antiferromagnetic Skyrmions," *Adv. Funct. Mater.*, vol. 33, no. 49, p. 2303133, Dec. 2023, doi: 10.1002/ADFM.202303133.